\documentclass[nofootinbib,a4paper,aps,twocolumn,prd,10pt,superscriptaddress]{revtex4}

\usepackage{graphicx}
\usepackage{amsmath}
\usepackage{amsfonts}
\usepackage{amsthm}
\usepackage{mathrsfs}
\usepackage{amssymb}
\usepackage{epsfig}
\usepackage{amscd}
\usepackage{dcolumn}
\usepackage{bm}
\usepackage{natbib}
\usepackage{url}
\usepackage{xspace}
\usepackage{kantlipsum}
\usepackage{setspace}
\usepackage{comment}

\usepackage[normalem]{ulem}
\usepackage{appendix}
\usepackage{algorithm}
\usepackage{algorithmic}

\usepackage{siunitx}
\usepackage{paralist}

\usepackage{placeins}

\usepackage{xcolor}

\usepackage{comment}

\usepackage{hyperref}
\hypersetup{%
    ,urlcolor=blue
    ,citecolor=blue
    ,linkcolor=blue
    }

\definecolor{gs}{RGB}{0,50,200}

\def\be{\begin{equation}}
\def\ee{\end{equation}}
\def\bea{\begin{eqnarray}}
\def\eea{\end{eqnarray}}

\begin{document}


\title{QPO constraints on the rotating Fan–Wang spacetime}

\author{Manas~\surname{Khassanov}}
\email[]{manas.khassanov@gmail.com}
\affiliation{Al-Farabi Kazakh National University, Al-Farabi Ave., 71, Almaty, 050040, Kazakhstan}

\author{Gulnara~\surname{Suliyeva}}
\email[]{g.suliyeva@mail.ru}
\affiliation{Al-Farabi Kazakh National University, Al-Farabi Ave., 71, Almaty, 050040, Kazakhstan}
\affiliation{Fesenkov Astrophysical Institute, Observatory 23, 050020 Almaty, Kazakhstan}

\author{Marco~\surname{Muccino}}
\email[]{marco.muccino@unicam.it}
\affiliation{Universit\`a di Camerino, Via Madonna delle Carceri 9, 62032 Camerino, Italy}
\affiliation{Al-Farabi Kazakh National University, Al-Farabi Ave., 71, Almaty, 050040, Kazakhstan}
\affiliation{INAF, Catania Astrophysical Observatory, Via S.Sofia 79, 95123 Catania, Italy}
\affiliation{ICRANet, Piazza della Repubblica 10, 65122 Pescara, Italy}

\author{Kuantay~\surname{Boshkayev}}
\email[]{kuantay@mail.ru}
\affiliation{Al-Farabi Kazakh National University, Al-Farabi Ave., 71, Almaty, 050040, Kazakhstan}
\affiliation{Kazakh-British Technical University, Tole Bi Str.\ 59, 050000 Almaty, Kazakhstan}

\author{Daniya~\surname{Utepova}}
\email[]{utepova\_daniya@mail.ru}
\affiliation{Abai Kazakh National Pedagogical University, Dostyk Ave., 13, Almaty, 050010, Kazakhstan}
\affiliation{Al-Farabi Kazakh National University, Al-Farabi Ave., 71, Almaty, 050040, Kazakhstan}

\author{Ruslan~\surname{Spassyuk}}
\email[]{ruslan.spassyuk10@gmail.com}
\affiliation{Al-Farabi Kazakh National University, Al-Farabi Ave., 71, Almaty, 050040, Kazakhstan}
\affiliation{Fesenkov Astrophysical Institute, Observatory 23, 050020 Almaty, Kazakhstan}

\date{\today}

\begin{abstract}
In this work, we investigate a rotating regular black hole described by the Fan–Wang solution, which arises from nonlinear electrodynamics coupled to gravity. We examine the motion of neutral test particles along circular equatorial orbits and derive the corresponding orbital parameters, including the angular velocity, specific angular momentum, and specific energy. By perturbing these orbits in the radial and vertical directions, we calculate the associated fundamental epicyclic frequencies. We then employ the relativistic precession model, to interpret the observed quasiperiodic oscillations (QPOs) in terms of these frequencies. We numerically determine the radii of the innermost stable circular orbits and perform Markov chain Monte Carlo simulations to constrain the spacetime parameters. Using observational and theoretical estimates of the black hole mass and spin parameter from the literature, together with theoretically motivated priors on the Fan–Wang parameter, we obtain the best-fit parameter values. The inferred black hole mass, spin parameter, and QPO emission radii are consistent with the corresponding estimates reported in the literature. Furthermore, the dimensionless Fan-Wang parameter satisfies $l/M < 8/27 \approx 0.296$, in agreement with its theoretical upper bound.
\end{abstract}

\maketitle 
\section{Introduction}
The extremely strong gravitational fields of black holes (BHs) make these objects particularly important for testing general relativity (GR). One of the most compelling pieces of evidence for the existence of such objects was the detection of gravitational waves produced by the merger of compact objects \cite{ligo2016tests}. Another important observational achievement was the imaging of M87$^*$ \cite{2019ApJ...875L...1E} and Sgr A$^*$ \cite{2022ApJ...930L..12E} by the Event Horizon Telescope (EHT) Collaboration.

Despite the success of GR in describing a broad range of astrophysical phenomena, the Schwarzschild \cite{1916AbhKP1916..189S}, Reissner–Nordström \cite{1916AnP...355..106R,1918KNAB...20.1238N}, and Kerr \cite{PhysRevLett.Kerr} solutions contain curvature singularities. Under general conditions such singularities are unavoidable in
gravitational collapse \cite{1965PhRvL..14...57P,1970RSPSA.314..529H}, and they are
widely regarded as a sign that the classical theory breaks down rather than
as a physical feature.

Regular BHs (RBHs) provide a possible way of avoiding this problem. Such solutions are frequently obtained by coupling gravity to nonlinear electrodynamics (NED) and are characterized by finite curvature invariants. The first example was proposed by Bardeen \cite{1968Bardeen87B}. It was later shown to be an exact solution of GR coupled to a nonlinear magnetic monopole \cite{2000PhLB..493..149A}, and regular solutions with electric charge \cite{1998PhRvL..80.5056A} and general NED constructions of regular magnetic BHs \cite{2000PhRvL..85.4641B} were subsequently found. Well-known examples include the Dymnikova \cite{Dymnikova_GR_&_Gravitation235D}, Hayward \cite{hayward2006formation}, and Fan–Wang \cite{Wanfan2016construction} solutions; recent reviews can be found in Refs. \cite{maeda2022quest,bronnikov2023regular}. 

Within this class, the Fan–Wang solution provides a useful framework for studying the influence of a NED charge on BH geometry. Its static form is asymptotically flat, possesses a regular de Sitter-like core, and reduces to the Schwarzschild solution when the charge parameter vanishes\cite{suliyeva2025FW}. To incorporate rotation, we revisit a stationary and axisymmetric extension of this geometry obtained using the modified Newman–Janis procedure developed by Azreg-Aïnou \cite{azreg2014a,azreg2014b,azreg2014c}. This approach avoids the ambiguous complexification of the radial coordinate encountered in the standard Newman–Janis algorithm \cite{NewmanJanisProc1965} and allows the resulting metric to be written in Boyer–Lindquist coordinates. The rotating solution recovers the static Fan–Wang spacetime in the limit ($a\to0$) and reduces to the Kerr geometry when the charge parameter ($l$) vanishes.

It is worth noting that more general RBH solutions have been investigated in the literature, including the construction of their rotating counterparts and the derivation of the corresponding fundamental frequencies \cite{2017PhRvD..95h4037T}. These general solutions encompass several well-known regular BH models, including the Bardeen, Hayward, and Fan–Wang solutions, as special cases. The treatment in \cite{2017PhRvD..95h4037T} is primarily a theoretical analysis of the geodesic and epicyclic structure of the rotating RBH family. Since the relevant expressions have been derived in a general form, the present work focuses specifically on the rotating Fan–Wang spacetime and investigates its implications for quasi-periodic oscillations (QPOs).

Recent studies have investigated the shadow, horizon structure, and ergoregion of the rotating Fan–Wang BH, examining the effects of spin and the deviation parameter ($l$). Using shadow observables and EHT observations of Sgr A* and M87*, these studies constrained the allowed parameter space and found that the model is consistent with current observations. They also showed that increasing ($l$) reduces the shadow size and suppresses high-energy emission \cite{2026PDU....5102203G}.

At present, there is no direct observational evidence that compact objects carry a magnetic monopole charge. Nevertheless, from the theoretical perspective, its possible magnitude can be constrained indirectly by examining its influence on observable phenomena. However, horizon scale images from the EHT have already been used to bound charge-like parameters and deviations from the Kerr metric
\cite{2021PhRvD.103j4047K,2023CQGra..40p5007V}; X-ray timing offers a complementary,
independent constraint that is sensitive to the dynamics of matter rather
than to photon orbits.

In the Fan–Wang solution, the magnetic charge parameter $l$ enters the metric functions and thereby affects the motion of matter near the BH. This effect is reflected, in particular, in the properties of circular orbits and the associated fundamental frequencies. Comparing the theoretically predicted frequencies with astrophysical data provides a means of testing the Fan–Wang geometry and constraining the allowed range of its parameters.

QPOs observed in the X-ray emission from accreting compact objects provide a potential probe of orbital dynamics in the strong-field regime \cite{abramowicz2013}. They were established as a general feature of accreting compact objects
in X-ray timing studies \cite{2006csxs.book...39V,2006ARA&A..44...49R}. In
BH X-ray binaries the high-frequency (HF) QPOs have frequencies of
order $10^2$\,Hz, are often observed in pairs with commensurate frequencies
(close to a $3{:}2$ ratio), and are believed to originate in the innermost
disk \cite{2006ARA&A..44...49R,2019NewAR..8501524I}. Twin HF QPOs have been
reported in GRO\,J1655$-$40 \cite{2001ApJ...552L..49S}, XTE\,J1550$-$564
\cite{2002ApJ...564..962R}, GRS\,1915$+$105 and H1743$-$322 \cite{2006ApJ...637.1002R},
and a $3{:}2$ pair was also found in the intermediate-mass BH
candidate M82~X-1 \cite{2014Natur.513...74P}, consistent with the inverse scaling of
the frequencies with mass.

The QPOs' measured frequencies have been used both to estimate BH parameters and to test different spacetime models \cite{stella1999,rezzolla2003new,stuchlik2013multi}. Geodesic interpretations of QPOs include the relativistic precession model (RPM) \cite{stella1999,1998ApJ...492L..59S},
epicyclic-resonance models \cite{2001AcPPB..32.3605K,2001A&A...374L..19A,2004ApJ...609L..63A} and
diskoseismology \cite{1999PhR...311..259W}. Particularly, in the RPM the upper and lower HF QPOs
correspond to the Keplerian and periastron precession frequencies, respectively, and the nodal precession frequency corresponds to a low-frequency QPO. When a QPO triplet is available, the RPM yields
BH masses and spins that agree with independent measurements
\cite{2014MNRAS.437.2554M,2014MNRAS.439L..65M,2014MNRAS.444.2065I,2022MNRAS.517.1469M,2016AN....337..398M,2017MNRAS.467..145F}. In several theoretical models, HF QPOs are associated with orbital and epicyclic motion in the inner regions of accretion disks \cite{boshkayev2025constraints}. 

Since the Keplerian and epicyclic frequencies depend on the metric functions and their derivatives, both the rotation and magnetic charge parameters modify their predicted values. In particular, a nonzero magnetic charge parameter produces deviations from the Kerr limit. Comparing the calculated frequencies with observed QPOs can therefore place indirect constraints on the parameters of the rotating Fan–Wang spacetime.

In this paper we rederive the Keplerian, radial and vertical epicyclic frequencies of neutral test particles on circular equatorial orbits in the rotating Fan-Wang spacetime, for both co- and counter-rotating orbits. Within the RPM we then confront these frequencies with the twin-peak QPOs of several microquasars using
Bayesian inference with Markov chain Monte Carlo (MCMC) sampling, in order to
constrain the mass, spin, charge parameter $l$ and emission radius, and to
discuss the degeneracy with the Kerr spin.

Several aspects of the static Fan-Wang spacetime have already been explored. The motion of neutral test particles and classical effects of GR were analyzed in ~\cite{2023PhRvD.107h4003I,suliyeva2025FW}. The structure of accretion disks around the static Fan-Wang BH have been studied in
Refs.~\cite{2024PDU....4601566K}, and its optical appearance in the presence
of an anisotropic, quintessence-like exotic fluid was investigated in
~\cite{2026EPJP..141..436K}. All of these studies, however, are confined
to the non-rotating case. To our knowledge, the possible imprints on QPOs have not been examined previously. This is what motivates the present work: we perform numerical analyses on the observed QPO data of microquasars and derive spin and magnetic charge parameters in the rotating Fan-Wang spacetime.

This paper is organized as follows. In Sec.\ref{sec:fan_wang}, we introduce the static Fan--Wang spacetime and its rotating counterpart and discuss the corresponding horizon, extremality, and stationary limit structure. In Sec.\ref{sec:Orbits}, we derive the angular velocity, specific energy, and specific angular momentum of neutral massive particles moving along equatorial circular orbits. In Sec.\ref{sec:gravred}, we obtain the radial and vertical epicyclic frequencies, analyze the innermost stable circular orbit (ISCO) condition, and relate the fundamental frequencies to the observed QPOs within the relativistic precession model. In Sec.\ref{sec:defangle}, we present the MCMC analysis of four microquasars, constrain the spacetime parameters, and compare the inferred results with published Kerr estimates. Finally, Sec.\ref{sec:con} the main findings and outlines possible directions for further investigation.
 

\section{Fan--Wang spacetime} \label{sec:fan_wang} RBH solutions provide an important framework for investigating possible nonsingular alternatives to classical BH geometries. In this work, we consider the Fan--Wang spacetime, which arises in Einstein gravity coupled to NED and is characterized by the gravitational mass $M$ and an additional charge parameter $l$. The corresponding static solution is asymptotically flat, reduces to the Schwarzschild geometry when ($l=0$), and, for the particular Fan--Wang model with ($\mu=3$) and ($\nu=1$) considered here, possesses a regular de Sitter-like core for ($l>0$) \cite{Wanfan2016construction,maeda2022quest}. We first summarize the main properties of the static Fan--Wang spacetime and then employ the modified Newman--Janis procedure proposed by Azreg-A{\"i}nou to construct its stationary and axisymmetric rotating counterpart.

\subsection{Static Fan-Wang spacetime}
\label{subsec:static_fw}
In the static and spherically symmetric case, the Fan–Wang spacetime is described in standard curvature coordinates by the line element \cite{Wanfan2016construction,maeda2022quest,2023IJTP...62..202L} 
\begin{equation}
\label{gen_ds}
ds^{2} = -A(r)dt^{2} + \frac{dr^{2}}{A(r)} + r^{2}(d\theta^2 + \sin^{2}{\theta} d\phi^{2})
\end{equation}
Following the Damour–Solodukhin prescription \cite{dutta2022generalized}, the lapse function for a static RBH can be expressed in terms of a radial mass function as
\begin{equation}
\label{laps_func}
A(r) = 1-2\mathcal{M}(r)/r.
\end{equation}
For the particular member of the Fan--Wang family with ($\mu=3$) and ($\nu=1$) adopted in this work, the mass function is given by \cite{Wanfan2016construction,maeda2022quest}
\begin{equation}
\label{mass_func}
\mathcal{M}(r)=\frac{Mr^{3}}{(r + l)^{3}}.
\end{equation}
Consequently, the lapse function takes the explicit form
\begin{equation}
\label{fw_lapse}
A(r) = 1-\frac{2Mr^{2}}{(r + l)^{3}}.
\end{equation}

Here, $M > 0$ denotes the Arnowitt–Deser–Misner mass of the gravitational source, while $l \geq 0$ is the charge parameter associated with the nonlinear electromagnetic field. At large distances, $\mathcal{M}(r)$ approaches $M$, confirming the asymptotic flatness of the spacetime. For ($l=0$), Eq. (\ref{fw_lapse}) reduces to $A(r) = 1-2M/r.$, and the Schwarzschild solution is recovered.

For ($l>0$), the particular ($\mu=3$, $\nu=1$) Fan--Wang solution considered here is regular at ($r=0$) and possesses a de Sitter-like core \cite{Wanfan2016construction,maeda2022quest}. The horizons are determined by the condition ($A(r)=0$), which yields
\begin{equation}
\label{condition}
(r + l)^3-2Mr^2 = 0.
\end{equation}

For $0 < l < 8M/27$, this equation admits two positive roots corresponding to the inner and outer horizons. The two horizons merge in the extremal case,
\begin{equation}
l_{\mathrm{ext}}=\frac{8M}{27},
\qquad
r_{\mathrm{ext}}=\frac{16M}{27}
\end{equation}
whereas for $l > l_{\mathrm{ext}}$ no event horizon exists. We therefore restrict our analysis to the BH domain
\begin{equation}
0 \leq l/M \leq 8/27.
\end{equation}
The static geometry described above will be used as the seed spacetime for constructing its rotating counterpart.

\subsection{Rotating Fan--Wang spacetime}
\label{subsec:rotating_fw}

Here we review the construction of the rotating counterpart of the static Fan–Wang spacetime derived in  term of the modified Newman–Janis procedure developed by Azreg-Aïnou \cite{azreg2014a,azreg2014b,azreg2014c}. This method avoids the ambiguous complexification of the radial coordinate and allows the resulting stationary and axisymmetric geometry to be written in Boyer–Lindquist coordinates. 
The procedure starts from a general static and spherically symmetric seed metric, whose line element can be written as
\begin{equation} \label{general_seed_metric} ds^{2} = -G(r)\,dt^{2} +\frac{dr^{2}}{F(r)} +H(r)\left( d\theta^{2} +\sin^{2}\theta\,d\phi^{2} \right), \end{equation} 
where $G(r)$, $F(r)$, and $H(r)$ are radial metric functions \cite{azreg2014a, azreg2014b}. Comparing the general line element~(\ref{general_seed_metric}) with the static Fan--Wang metric~(\ref{gen_ds}), we identify 
\begin{equation} 
\label{fw_seed_functions} G(r)=F(r)=A(r)\qquad,\qquad H(r)=r^{2}. 
\end{equation}
The auxiliary function $K(r)$, introduced in the rotating construction, is defined as
\begin{equation} 
\label{aux_func}
K(r)=H(r)\sqrt{\frac{F(r)}{G(r)}}.
\end{equation}
Thus, for the Fan–Wang seed metric is $K(r)=r^{2}$. The corresponding differential constraints imposed on the rotating geometry admit the simple solution \cite{azreg2014a,azreg2014c}
\begin{equation} 
\label{Psi_func}
\Psi(r,\theta,a)=r^{2}+a^{2}{\cos^{2}{\theta}}
\end{equation}
which, for $a=0$, yields $\Psi(r,\theta,0)=H(r)=r^{2}$.

To express the resulting rotating geometry in a compact Kerr-like form, we introduce the standard quantity \cite{azreg2014a, azreg2014b}
\begin{equation}
\label{rho}
\rho^{2}
\equiv K(r)+a^{2}\cos^{2}\theta
= r^{2}+a^{2}\cos^{2}\theta.
\end{equation}
Using $K(r)=r^{2}$, the quantity $\rho^{2}$ coincides with the obtained function $\Psi(r,\theta,a)$, and hence the 
conformal factor $\Psi/\rho^{2}$ in the general rotating metric 
is unity. 
To complete the construction, the remaining functions entering the 
rotating line element are defined as
\begin{subequations}
\label{rot_functions}
\begin{align}
2f(r) &= K(r)-F(r)H(r),\\
\Delta(r) &= F(r)H(r)+a^{2},\\
\Sigma(r,\theta)&=\left[K(r)+a^{2}\right]^{2}-a^{2}\Delta(r)\sin^{2}\theta.
\end{align}
\end{subequations}
Here, \(f(r)\) is an auxiliary radial function associated with the mass distribution of the rotating spacetime and should not be confused with the static lapse function \(A(r)\). The function \(\Delta(r)\) determines the radial component of the metric and governs the location of the horizons, whereas \(\Sigma(r,\theta)\) enters the azimuthal metric component. 

Substituting $F(r)\!=\!A(r)$, $H(r)\!=\!K(r)\!=\!r^{2}$, together with the Fan–Wang lapse function, Eqs.~\eqref{rot_functions} reduce to
\begin{subequations}
\label{fw_rotating_functions}
\begin{align}
2f(r)
&=r^{2}\left[1-A(r)\right]
=\frac{2Mr^{4}}{(r+l)^{3}}
=2r\mathcal{M}(r),\\
\Delta(r)
&=r^{2}A(r)+a^{2}
=r^{2}+a^{2}-\frac{2Mr^{4}}{(r+l)^{3}},\\
\Sigma(r,\theta)
&=\left(r^{2}+a^{2}\right)^{2}
-a^{2}\Delta(r)\sin^{2}\theta.
\end{align}
\end{subequations}
Substituting these functions into the general Kerr-like line element yields the rotating Fan–Wang metric in Boyer–Lindquist coordinates:
\begin{equation}
\label{rotating_FW_metric}
\begin{aligned}
ds^{2}={}&
-\left(1-\frac{2Mr^{4}}
{(r+l)^{3}\rho^{2}}\right)dt^{2}
+\frac{\rho^{2}}{\Delta(r)}dr^{2}
+\rho^{2}d\theta^{2}
\\
&+\frac{\Sigma(r,\theta)\sin^{2}\theta}
{\rho^{2}}d\phi^{2}
-\frac{4aMr^{4}\sin^{2}\theta}
{(r+l)^{3}\rho^{2}}\,dt\,d\phi .
\end{aligned}
\end{equation}
The off-diagonal term $\propto dt\,d\phi$ describes the frame-dragging effect produced by the spacetime rotation.

The consistency of the resulting metric can be verified through two limiting cases. In the non rotating limit \(a\to0\), one has
\begin{equation}
\rho^{2}\to r^{2}, \qquad
\Delta(r)\to r^{2}A(r), \qquad
g_{t\phi}\to 0.
\end{equation}
and the rotating metric reduces to the original static Fan–Wang metric. In the limit \(l\to0\),
\begin{equation}
\mathcal{M}(r)\to M\ ,\ 
f(r)\to Mr\ ,\ 
\Delta(r)\to r^{2}-2Mr+a^{2}.
\end{equation}
and the Kerr geometry is recovered \cite{PhysRevLett.Kerr}.

The horizons of the rotating Fan--Wang spacetime are determined by
the real positive roots of
\(\Delta(r)=0\), or explicitly,
\begin{equation}
\label{fw_horizon_equation}
\left(r^{2}+a^{2}\right)(r+l)^{3}
-2Mr^{4}=0.
\end{equation}
For a subextremal BH, Eq.~\eqref{fw_horizon_equation}
has two positive roots
\begin{equation}
r_{-}<r_{+},
\end{equation}
which correspond to the inner and outer horizons, respectively.
The largest positive root \(r_{+}\) is identified with the outer
event horizon.

When the inner and outer horizons coincide, the BH becomes
extremal. The degenerate horizon \(r_{\mathrm e}\) then satisfies
\begin{equation}
\Delta(r_{\mathrm e})=0
\qquad,\qquad
\Delta'(r_{\mathrm e})=0.
\end{equation}
Solving these conditions simultaneously gives
\begin{equation}
\label{fw_extremal_relations}
M=
\frac{(r_{\mathrm e}+l)^{4}}
{r_{\mathrm e}^{2}(r_{\mathrm e}+4l)}
\qquad,\qquad
a_{\mathrm{ext}}^{2}
=
\frac{r_{\mathrm e}^{2}(r_{\mathrm e}-2l)}
{r_{\mathrm e}+4l}.
\end{equation}
These relations determine the extremal boundary in the parameter
space of the rotating solution. At the nonrotating endpoint of the
extremal curve, \(a_{\mathrm{ext}}=0\), they yield
\begin{equation}
l_{\mathrm{ext}}=\frac{8M}{27}
\qquad,\qquad
r_{\mathrm e}=\frac{16M}{27}.
\end{equation}
In the limit \(l\to0\), Eq.~(\ref{fw_extremal_relations}) reduces to
the extremal Kerr relations
\begin{equation}
r_{\mathrm e}=M
\qquad,\qquad
a_{\mathrm{ext}}=M.
\end{equation}

In addition to the horizons, the rotating spacetime possesses
stationary-limit surfaces determined by the condition
\(g_{tt}=0\). From Eq.~(\ref{rotating_FW_metric}), the corresponding
metric component is
\begin{equation}
g_{tt}
=
-\left[
1-\frac{2Mr^{4}}
{(r+l)^{3}
\left(r^{2}+a^{2}\cos^{2}\theta\right)}
\right].
\end{equation}
Consequently, the stationary-limit surfaces satisfy
\begin{equation}
\label{fw_ergosurface_equation}
\left(r^{2}+a^{2}\cos^{2}\theta\right)(r+l)^{3}
-2Mr^{4}=0.
\end{equation}
In general, Eq.~(\ref{fw_ergosurface_equation}) must be solved
numerically for each polar angle. We denote its inner and outer
roots by \(r_{\mathrm E}^{-}(\theta)\) and
\(r_{\mathrm E}^{+}(\theta)\), respectively.

At the symmetry axis, \(\theta=0,\pi\), the stationary-limit
equation coincides with the horizon equation. Therefore,
\begin{equation}
r_{\mathrm E}^{\pm}(0)
=
r_{\mathrm E}^{\pm}(\pi)
=
r_{\pm}.
\end{equation}
Thus, the inner and outer stationary-limit surfaces meet the
corresponding horizons at the poles.

In the equatorial plane, \(\theta=\pi/2\),
Eq.~(\ref{fw_ergosurface_equation}) reduces to
\begin{equation}
r^{2}\left[(r+l)^{3}-2Mr^{2}\right]=0.
\end{equation}
Hence, the nonzero equatorial stationary-limit radii coincide with
the horizon radii of the static Fan--Wang spacetime and are
independent of the rotation parameter \(a\). In particular,
\begin{equation}
r_{\mathrm E}^{\pm}\left(\frac{\pi}{2}\right)
=
r_{\mathrm{static}}^{\pm}.
\end{equation}
The exterior ergoregion is located between the outer event horizon
and the outer stationary-limit surface,
\begin{equation}
r_{+}<r<r_{\mathrm E}^{+}(\theta).
\end{equation}

\begin{figure*}[!t]
    \centering

    \includegraphics[width=0.45\textwidth]
    {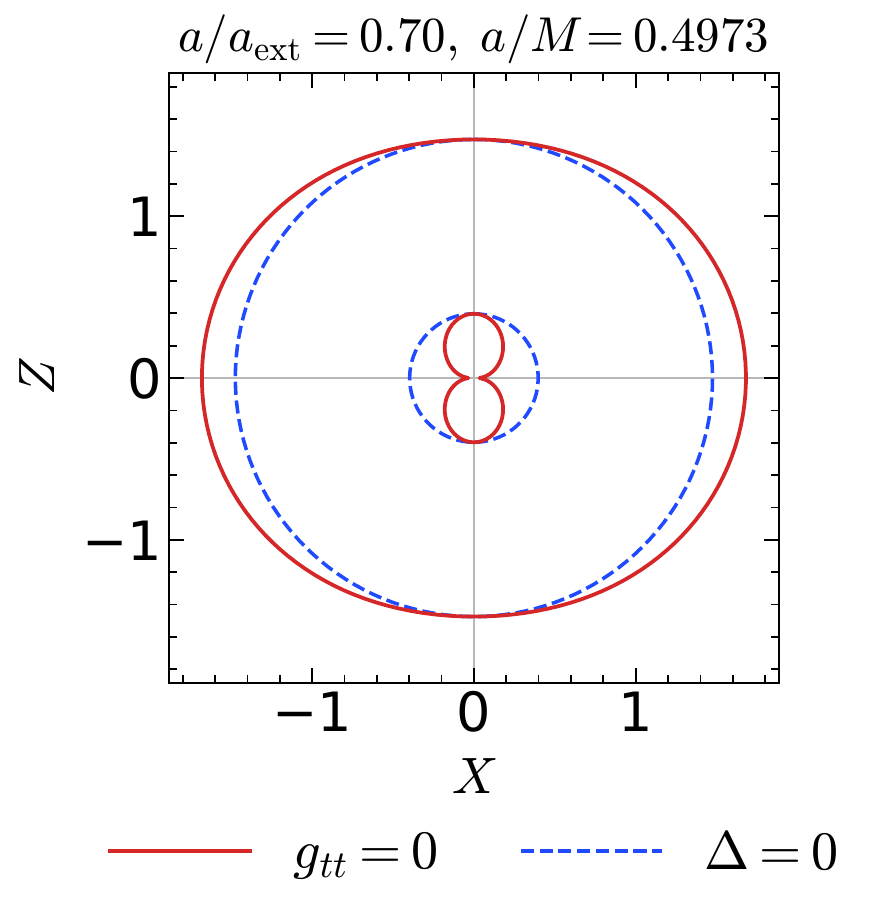}
    \hfill
    \includegraphics[width=0.45\textwidth]
    {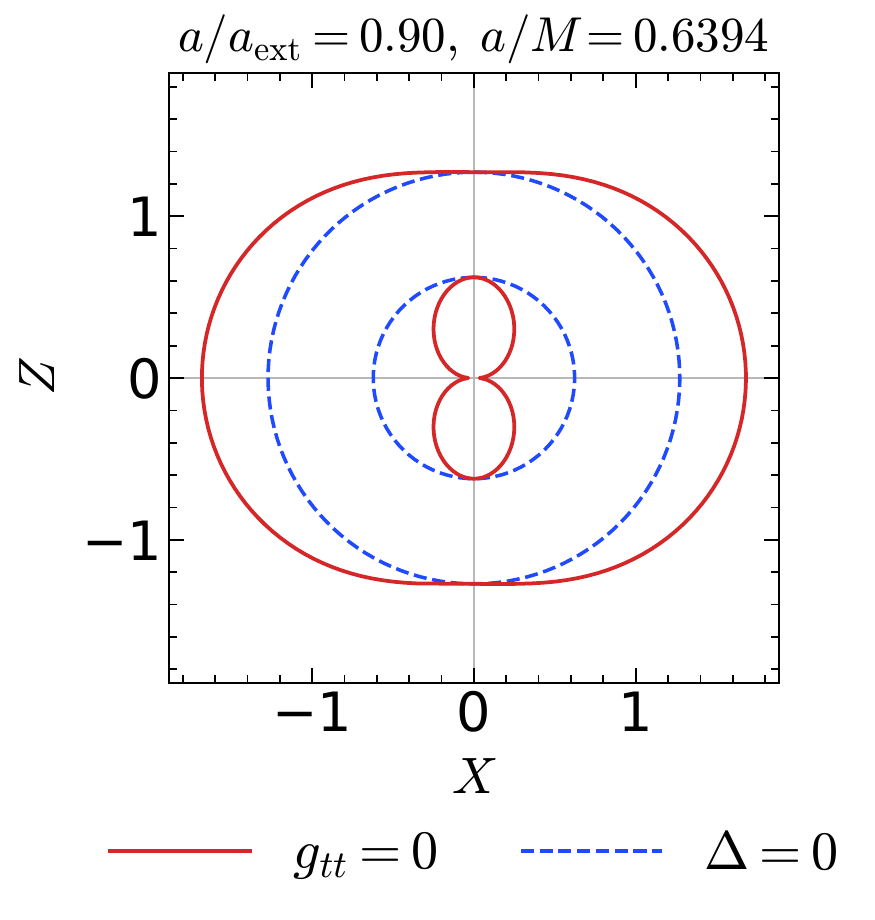}

    \vspace{0.8em}

    \includegraphics[width=0.45\textwidth]
    {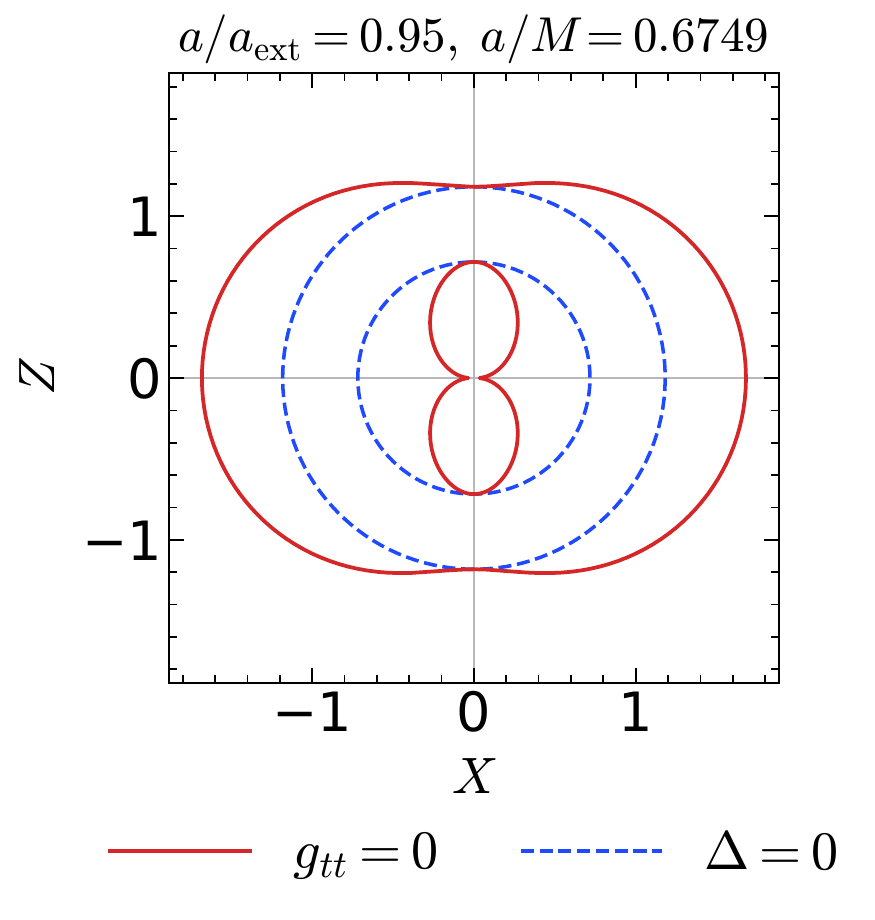}
    \hfill
    \includegraphics[width=0.45\textwidth]
    {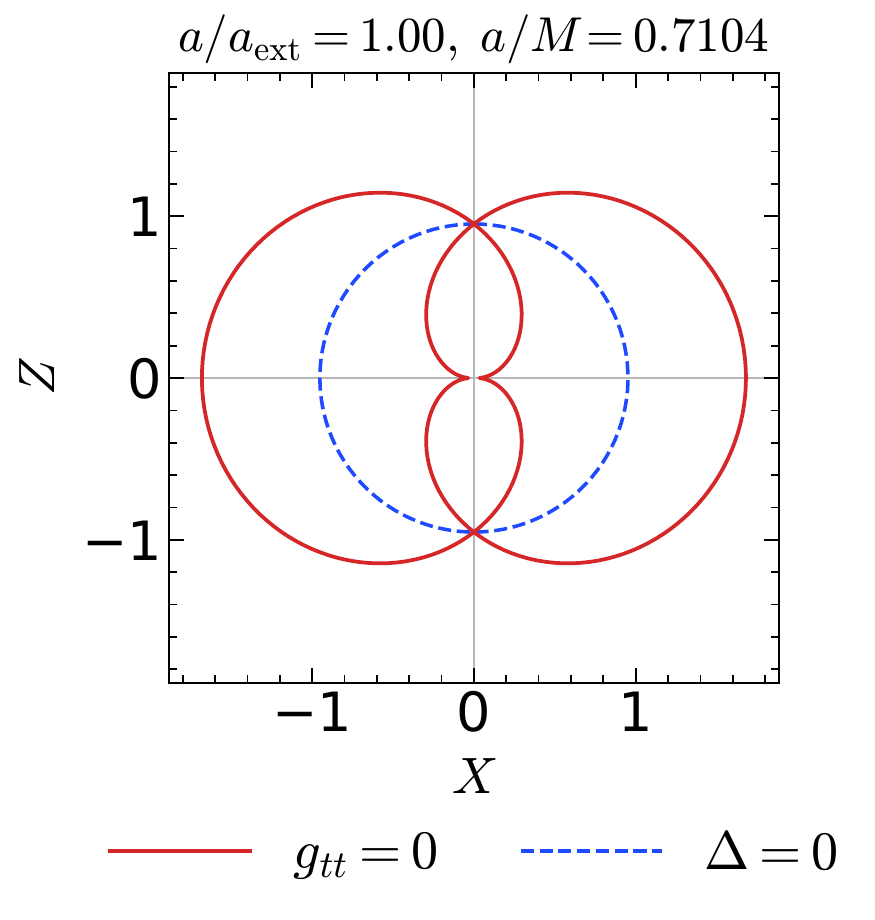}

    \caption{
    Coordinate meridional representation of the inner and outer
    horizons (blue dashed curves) and the inner and outer
    stationary-limit surfaces (red solid curves) of the rotating
    Fan--Wang spacetime for \(M=1\) and \(l/M=0.10\), corresponding to
    \(a_{\mathrm{ext}}/M\simeq0.7104\).
    The panels correspond to
    (a) \(a/a_{\mathrm{ext}}=0.70\),
    (b) \(a/a_{\mathrm{ext}}=0.90\),
    (c) \(a/a_{\mathrm{ext}}=0.95\), and
    (d) \(a/a_{\mathrm{ext}}=1.00\).
    At the poles, each stationary-limit surface coincides with the
    corresponding horizon. In the extremal case
    \(a/a_{\mathrm{ext}}=1\), the inner and outer horizons merge.
    The exterior ergoregion lies between the outer horizon and the
    outer stationary-limit surface. The curves are displayed using
    \(X=r\sin\theta\) and \(Z=r\cos\theta\).
    }
    \label{fig:fw_horizons_ergosurfaces}
\end{figure*}

To illustrate the effects of rotation, in
Fig.~\ref{fig:fw_horizons_ergosurfaces} we set
\(M=1\) and \(l/M=0.10\). For these parameters,
Eq.~(\ref{fw_extremal_relations}) gives
\begin{equation}
\frac{r_{\mathrm e}}{M}\simeq0.95245
\qquad,\qquad
\frac{a_{\mathrm{ext}}}{M}\simeq0.71043.
\end{equation}
The different panels correspond to
\(a/a_{\mathrm{ext}}=0.70\), \(0.90\), \(0.95\), and \(1.00\).

In the following analysis, we consider only parameter combinations
for which an outer event horizon exists and restrict the motion of
test particles to the region \(r>r_{+}\).

\section{Circular orbits in the equatorial plane}
\label{sec:Orbits}
We consider neutral massive test particles moving along timelike geodesics in the rotating Fan–Wang spacetime. Since the metric is independent of the coordinates \(t\) and \(\phi\), the specific energy \(E=-u_t\) and the azimuthal specific angular momentum \(L=u_\phi\) are conserved. Solving these relations for the corresponding contravariant components of the four-velocity gives
\begin{equation}
u^{t}=\frac{E g_{\phi\phi}+L g_{t\phi}}{g_{t\phi}^{2}-g_{tt}g_{\phi\phi}}
\qquad,\qquad
u^{\phi}=-\frac{E g_{t\phi}+L g_{tt}}{g_{t\phi}^{2}-g_{tt}g_{\phi\phi}}.
\end{equation}
Substituting these expressions into the normalization condition \(u^\mu u_\mu=-1\), the radial and vertical motion can be represented as \(g_{rr}\dot r^{\,2}+g_{\theta\theta}\dot\theta^{\,2}=V_{\mathrm{eff}}\), where the effective potential is given by
\begin{equation}
\label{eff_pot}
V_{\mathrm{eff}}
=
\frac{
E^{2}g_{\phi\phi}
+2ELg_{t\phi}
+L^{2}g_{tt}
}{
g_{t\phi}^{2}-g_{tt}g_{\phi\phi}
}
-1.
\end{equation}
A circular orbit of radius \(r_0\) in the equatorial plane is characterized by \(\theta=\pi/2\) and \(\dot r=\dot\theta=0\). Therefore, the circular-orbit conditions are \(V_{\mathrm{eff}}=0\) and \(\partial_rV_{\mathrm{eff}}=0\), evaluated at \(r=r_0\) and \(\theta=\pi/2\). Owing to the reflection symmetry about the equatorial plane, \(\partial_\theta V_{\mathrm{eff}}=0\) is automatically satisfied at \(\theta=\pi/2\).

The Keplerian angular velocity with respect to the coordinate time \(t\) is defined as \(\Omega_\phi=d\phi/dt=u^\phi/u^t\). Eliminating \(E\) and \(L\) by means of the circular-orbit conditions gives the general expression
\begin{equation}
\Omega_{\phi}^{\pm}
=
\frac{
-\partial_{r}g_{t\phi}
\pm
\sqrt{
\left(\partial_{r}g_{t\phi}\right)^{2}
-\left(\partial_{r}g_{tt}\right)
 \left(\partial_{r}g_{\phi\phi}\right)
}
}{
\partial_{r}g_{\phi\phi}
}.
\end{equation}
To express the orbital quantities in a compact form, we introduce
\begin{subequations}
\begin{align}
\omega(r)&=\frac{\sqrt{M(r-2l)}}{(r+l)^{2}},\\
Q^{(s)}(r)&=A(r)-r^{2}\omega^{2}(r)+2sa\omega(r).
\end{align}\end{subequations}
where \(s=+1\) and \(s=-1\) label the co-rotating and counter-rotating branches, respectively, for \(a>0\). Using the explicit Fan–Wang lapse function, the second auxiliary function can equivalently be written as
\begin{equation}
Q^{(s)}(r)
=
1-\frac{3Mr^{3}}{(r+l)^{4}}
+2sa\omega(r).
\end{equation}
For notational simplicity, the subscript \(0\) denoting the circular-orbit radius is omitted in the following expressions. Evaluating the general angular-velocity formula for the rotating Fan–Wang metric yields
\begin{equation}
\Omega_{\phi}^{(s)}
=
\frac{s\,\omega(r)}
{1+s a\omega(r)}.
\end{equation}
The corresponding specific energy, azimuthal specific angular momentum, and temporal component of the four-velocity are
\begin{subequations}
\label{circular_orbit_quantities}
\begin{align}
E^{(s)}
&=
\frac{A(r)+sa\omega(r)}
{\sqrt{Q^{(s)}(r)}},
\\
L^{(s)}
&=
\frac{s\left(r^{2}+a^{2}\right)\omega(r)
-a\left[1-A(r)\right]}
{\sqrt{Q^{(s)}(r)}},
\\
u^{(s),t}
&=
\frac{1+sa\omega(r)}
{\sqrt{Q^{(s)}(r)}}.
\end{align}
\end{subequations}
Physically admissible timelike circular orbits require \(\mathcal Q^{(s)}(r)>0\) and must lie outside the outer horizon \(r_+\). The reality of \(\omega(r)\) additionally requires \(r\geq 2l\). In the nonrotating limit \(a\to0\), these expressions reduce to the corresponding results for the static Fan–Wang spacetime, whereas the Kerr expressions are recovered when \(l\to0\).

\section{The frequencies of epicyclic motion}
\label{sec:gravred}

The stability of equatorial circular orbits can be investigated by considering sufficiently small radial $\delta r$ and vertical $\delta\theta$ displacements from an unperturbed orbit of radius \(r_0\), namely
\begin{equation}
r=r_{0}+\delta{r}
\qquad,\qquad
\theta=\frac{\pi}{2}+\delta{\theta}.
\end{equation}
Expanding the equations of motion to first order in $\delta r$ and $\delta\theta$ yields two decoupled harmonic oscillators \cite{kluzniak2005resonant}
\begin{equation}
\frac{d^{2}\delta r}{dt^{2}}
+[\Omega_{r}^{(s)}]^{2}\delta r=0
\qquad,\qquad
\frac{d^{2}\delta\theta}{dt^{2}}
+[\Omega_{\theta}^{(s)}]^{2}\delta\theta=0,
\end{equation}
where \(\Omega_{r}^{(s)}\) and \(\Omega_{\theta}^{(s)}\) are the radial and vertical epicyclic angular frequencies, respectively, measured with respect to the coordinate time \(t\). In terms of the effective potential of Eq.~\eqref{eff_pot}, these frequencies are given by
\begin{subequations}
\begin{align}
\Omega_{r}^{2}
&=
-\frac{1}{2g_{rr}(u^{t})^{2}}
\left.
\frac{\partial^{2}V_{\mathrm{eff}}}{\partial r^{2}}
\right|_{r=r_{0},\,\theta=\pi/2},
\\
\Omega_{\theta}^{2}
&=
-\frac{1}{2g_{\theta\theta}(u^{t})^{2}}
\left.
\frac{\partial^{2}V_{\mathrm{eff}}}{\partial\theta^{2}}
\right|_{r=r_{0},\,\theta=\pi/2}.
\end{align}
\end{subequations}
When evaluating these derivatives, \(E^{(s)}\) and \(L^{(s)}\) are held fixed at their circular-orbit values. 
Substituting the metric coefficients of the rotating Fan–Wang spacetime and the orbital quantities obtained in the previous section gives the radial epicyclic frequency
\begin{equation}
\label{radial_epicyclic_frequency}
\begin{aligned}
\left[\Omega_{r}^{(s)}\right]^{2}
={}&
\frac{1}{r^{2}\left[1+sa\omega(r)\right]^{2}}
\Bigg\{
\omega^{2}(r)
\\
&\hspace{-2.5em}\times
\Big[
3r^{2}A(r)-a^{2}
-4r^{4}\omega^{2}(r)+8sar^{2}\omega(r)
\Big]-
\\
&\frac{2M\left(r^{2}-4lr+l^{2}\right)}
{(r+l)^{5}} \left[a^{2}+r^{2}A(r)\right]
\Bigg\}.
\end{aligned}
\end{equation}
while the vertical epicyclic frequency is
\begin{equation}
\label{vertical_epicyclic_frequency}
\begin{aligned}
\left[\Omega_{\theta}^{(s)}\right]^{2}
={}&
\frac{1}{\left[1+sa\omega(r)\right]^{2}}
\Bigg\{
\omega^{2}(r)
\left(1+\frac{a^{2}}{r^{2}}\right)+
\\
&\frac{a^{2}\left[1-A(r)\right]}{r^{4}}
-\frac{2sa\left[1-A(r)\right]\omega(r)}{r^{2}}
\Bigg\}.
\end{aligned}
\end{equation}
The above expressions show that rotation lifts the degeneracy between the vertical and Keplerian frequencies. In the nonrotating limit \(a\to0\), spherical symmetry is restored and the frequencies satisfy $[\Omega_{\theta}^{(s)}]^{2}=[{\Omega_{\phi}^{(s)}}]^2=\omega^{2}(r)$. For \(a\neq0\), spherical symmetry is reduced to axial symmetry, and the vertical and azimuthal motions are no longer dynamically equivalent. Consequently, \([\Omega_{\theta}^{(s)}]^{2}\) generally differs from $[\Omega_\phi^{(s)}]^2$. The static Fan–Wang and Kerr expressions are recovered in the limits \(a\to0\) and \(l\to0\), respectively.

Finally, for stable circular orbits, the Keplerian, radial epicyclic, and vertical epicyclic frequencies are respectively defined as 
$f_{\phi}^{(s)}=|\Omega_{\phi}^{(s)}|/(2\pi)$, $f_{r}^{(s)}=\Omega_{r}^{(s)}/(2\pi)$, $f_{\theta}^{(s)}=\Omega_{\theta}^{(s)}/(2\pi)$ where the positive square roots of the squared epicyclic frequencies are taken. The absolute value in the definition of \(f_{\phi}^{(s)}\) ensures that the physical frequency remains positive for both co-rotating and counter-rotating orbits. Within the relativistic precession model, the upper high-frequency  QPO is identified with the Keplerian frequency, whereas the lower 
high-frequency QPO is associated with the periastron-precession 
frequency. Thus,
\begin{equation}
f_{U}^{(s)}=f_{\phi}^{(s)}
\qquad,\qquad
f_{L}^{(s)}=f_{\phi}^{(s)}-f_{r}^{(s)}.
\end{equation}
In a rotating spacetime, the difference between the Keplerian and 
vertical epicyclic frequencies also gives rise to nodal precession. 
Its frequency can be defined as
\begin{equation}
f_{\mathrm{nod}}^{(s)}
=
\left|
f_{\phi}^{(s)}-f_{\theta}^{(s)}
\right|.
\end{equation}
Physically, nodal precession describes the gradual rotation of the line of nodes of a slightly tilted orbit around the spin axis of the BH. It arises because the vertical oscillation is not completed in exactly the same period as the azimuthal revolution, an effect associated with frame dragging \cite{1998ApJ...492L..59S,mashhoon1984}. In the nonrotating limit \(a\to0\), spherical symmetry is restored, \(f_\theta^{(s)}=f_\phi^{(s)}\), and therefore \(f_{\mathrm{nod}}^{(s)}=0\).

In the full RPM, the nodal-precession frequency may be associated with a low-frequency type-C QPO, while \(f_L\) and \(f_U\) correspond to the lower and upper high-frequency QPOs, respectively. However, since the present analysis is based only on the observed twin HF QPOs, \(f_{\mathrm{nod}}^{(s)}\) is not included in the fitting procedure. It is introduced here to describe the complete set of characteristic precession frequencies supported by the rotating spacetime.

Another physical quantity of particular importance is the radius of the ISCO. Using the expressions for $E^{(s)}$ and $L^{(s)}$ given in Eq.~(\ref{circular_orbit_quantities}) and imposing $dE^{(s)}/dr=0$ or $dL^{(s)}/dr=0$ leads to an implicit equation for the ISCO radius $r_{\mathrm{ISCO}}^{(s)}$. Along the family of equatorial circular geodesics, these two conditions are equivalent and coincide with the vanishing of the radial epicyclic frequency, $[\Omega_{r}^{(s)}]^{2}$, given in Eq.~(\ref{radial_epicyclic_frequency}). Introducing the dimensionless quantities
\begin{equation}
x=\frac{r}{M},
\qquad
\lambda=\frac{l}{M},
\qquad
\chi=\frac{a}{M}.
\end{equation}
we define the dimensionless ISCO residual as
\begin{equation}
\label{isco_dimensionless}
\begin{aligned}
\mathcal I^{(s)}(x;\lambda,\chi)
={}&
(x+\lambda)^{3}
\left(x^{2}+5\lambda x-8\lambda^{2}\right)
-6x^{3}(x-\lambda)
\\
&+8s\chi(x+\lambda)^{2}(x-2\lambda)^{3/2}
\\
&-\frac{3\chi^{2}}{x}
(x-3\lambda)(x+\lambda)^{3}.
\end{aligned}
\end{equation}
The ISCO radius is therefore determined by
\begin{equation}
\mathcal I^{(s)}
\left(x_{\mathrm{ISCO}}^{(s)};\lambda,\chi\right)=0.
\end{equation}
For prescribed values of \(\lambda\), \(\chi\), and \(s\), this equation is solved numerically for \(x\). The physically relevant solution \(x_{\mathrm{ISCO}}^{(s)}\) lies outside the outer event horizon, satisfies $
x_{\mathrm{ISCO}}^{(s)}\geq 2\lambda, 
Q^{(s)}\!\left(x_{\mathrm{ISCO}}^{(s)}\right)>0,
$ and bounds the family of stable circular orbits connected continuously to spatial infinity. The corresponding dimensional ISCO radius is
\begin{equation}
r_{\mathrm{ISCO}}^{(s)}
=
M x_{\mathrm{ISCO}}^{(s)}.
\end{equation}
To illustrate the numerical determination of this radius, we introduce the normalized ISCO function
\begin{equation}
\widehat{\mathcal I}^{(s)}(x;\lambda,\chi)
=
\frac{\mathcal I^{(s)}(x;\lambda,\chi)}{x^3}.
\end{equation}
Since $x>0$ in the physically relevant region, this normalization does not change the locations of the positive roots. Consequently,
\begin{equation}
\widehat{\mathcal I}^{(s)}=0
\quad\Longleftrightarrow\quad
\mathcal I^{(s)}=0.
\end{equation}

\begin{figure*}[!t]
    \centering

    \includegraphics[width=0.45\textwidth]
    {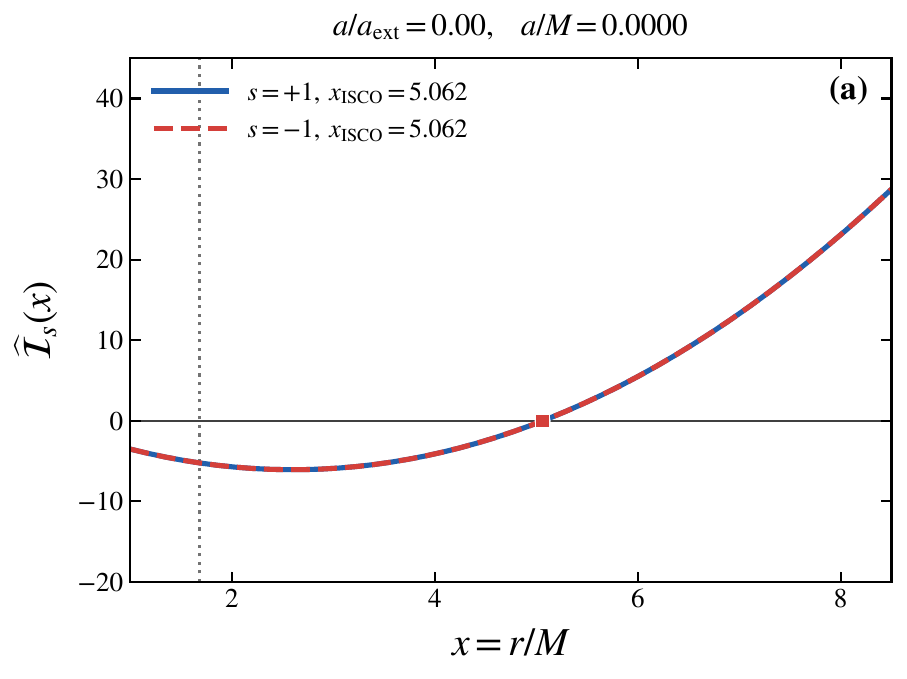}
    \hfill
    \includegraphics[width=0.45\textwidth]
    {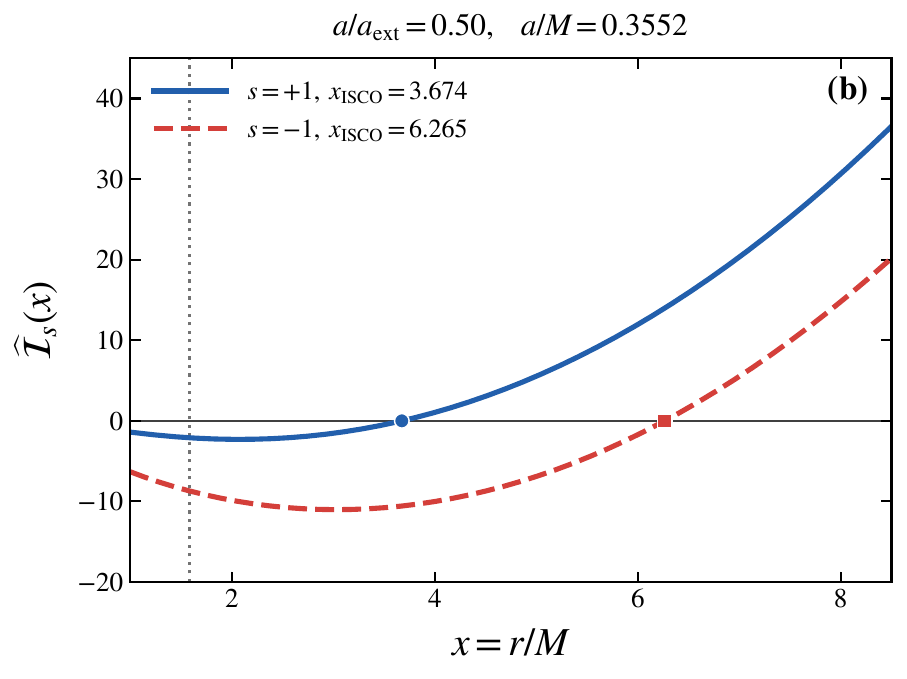}

    \par\vspace{0.2em}

    \includegraphics[width=0.45\textwidth]
    {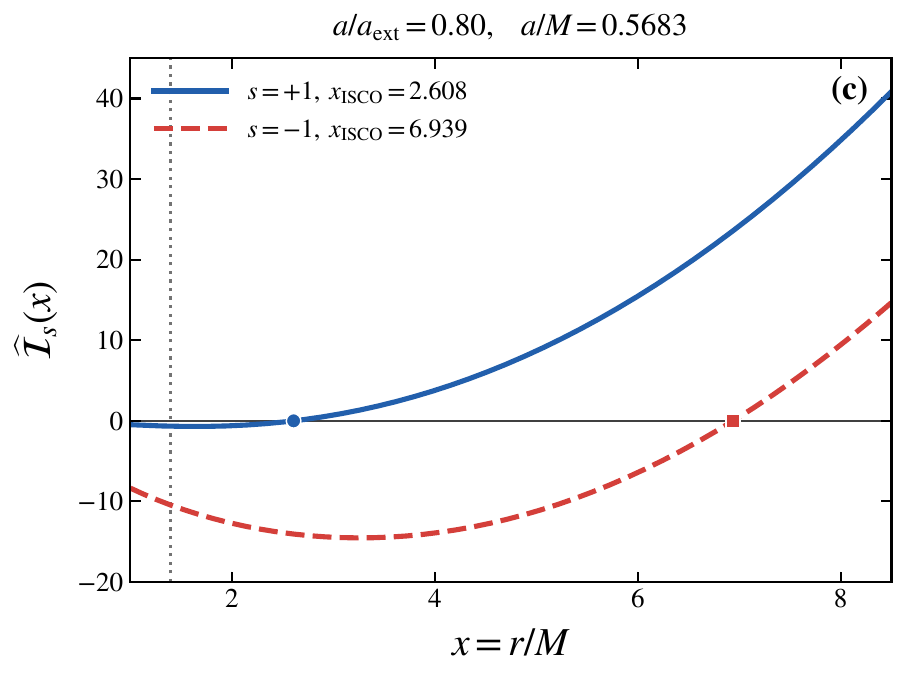}
    \hfill
    \includegraphics[width=0.45\textwidth]
    {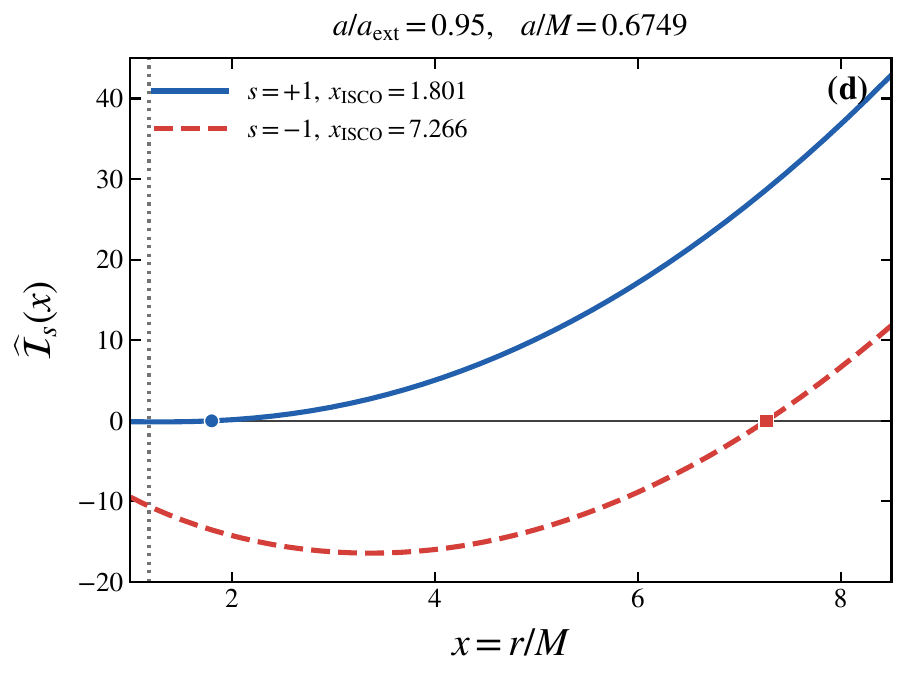}

    \caption{
    Normalized ISCO functions
    \(\widehat{\mathcal I}_{+}\) (blue solid curves) and
    \(\widehat{\mathcal I}_{-}\) (red dashed curves) of the rotating
    Fan--Wang spacetime for \(M=1\) and
    \(\lambda=l/M=0.10\). For this charge parameter,
    \(a_{\mathrm{ext}}/M=0.7104\).
    The panels correspond to
    (a) \(a/a_{\mathrm{ext}}=0\),
    (b) \(a/a_{\mathrm{ext}}=0.50\),
    (c) \(a/a_{\mathrm{ext}}=0.80\), and
    (d) \(a/a_{\mathrm{ext}}=0.95\).
    The marked intersections with the horizontal zero line identify
    the physical co-rotating and counter-rotating ISCO radii.
    The vertical dotted line indicates the position of the outer
    event horizon. The normalization by \(x^{3}\) does not alter
    the positive roots of the ISCO equation.
    }
    \label{fig:fw_isco_func}
\end{figure*}

As shown in Fig.~\ref{fig:fw_isco_func}, in the nonrotating case, the two orbital branches coincide and give
$x_{\mathrm{ISCO}}^{(+)}=x_{\mathrm{ISCO}}^{(-)}=5.062.$ Rotation separates the two branches: the co-rotating ISCO moves inward, whereas the counter-rotating ISCO moves outward. For example, at \(a/a_{\mathrm{ext}}=0.95\), the corresponding radii are $x_{\mathrm{ISCO}}^{(+)}=1.801, x_{\mathrm{ISCO}}^{(-)}=7.266$.

Although the ISCO equation cannot generally be inverted to obtain a compact closed-form expression for \(x_{\mathrm{ISCO}}^{(s)}\), it is quadratic in the dimensionless spin parameter \(\chi\). For convenience, we introduce
\begin{subequations}
\label{just_expression}
\begin{align}
P(x,\lambda)&=
(x+\lambda)^{3}\!
\left(x^{2}\!+\!5\lambda x\!-\!8\lambda^{2}\right)\!
-6x^{3}(x-\lambda),
\\
B(x,\lambda)&=
(x+\lambda)^{2}(x-2\lambda)^{3/2},
\\
C(x,\lambda)&=
\frac{(x-3\lambda)(x+\lambda)^{3}}{x}.
\end{align}
\end{subequations}

The ISCO residual can then be written compactly as
\begin{equation}
\mathcal I^{(s)}
=
P(x,\lambda)
+8s\chi B(x,\lambda)
-3\chi^2C(x,\lambda).
\end{equation}
Solving \(\mathcal I_s=0\) with respect to \(\chi\) and selecting the branch continuously connected to the nonrotating solution gives
\begin{equation}
\label{chi_isco}
\begin{aligned}
\chi&=\chi_{\mathrm{ISCO}}^{(s)}(x,\lambda) =
\frac{s}{6C(x,\lambda)}
\Bigg[
8B(x,\lambda)
\\
&\quad
-\sqrt{
64B^{2}(x,\lambda)
+12C(x,\lambda)P(x,\lambda)
}
\Bigg].
\end{aligned}
\end{equation}
This expression provides an analytical parametrization of the ISCO boundary: for specified \(x\), \(\lambda\), and orbital branch \(s\), it gives the corresponding value of the dimensionless rotation parameter. Conversely, when \(\lambda\), \(\chi\), and \(s\) are prescribed, the original equation \(\mathcal I^{(s)}=0\) must generally be solved numerically for \(x_{\mathrm{ISCO}}^{(s)}\).

\begin{figure*}[!t]
    \centering

    \includegraphics[width=0.48\textwidth]
    {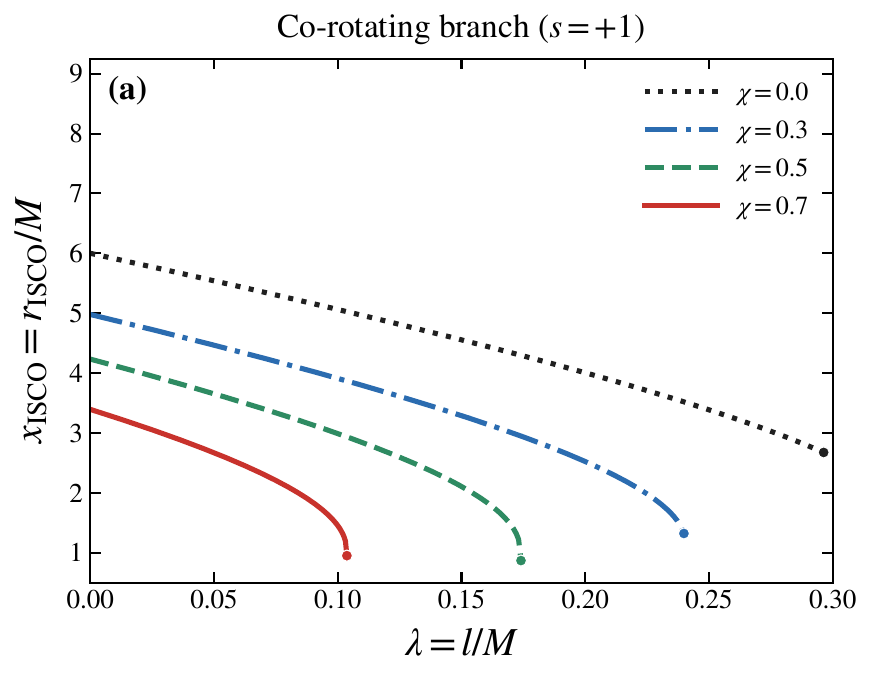}
    \hfill
    \includegraphics[width=0.48\textwidth]
    {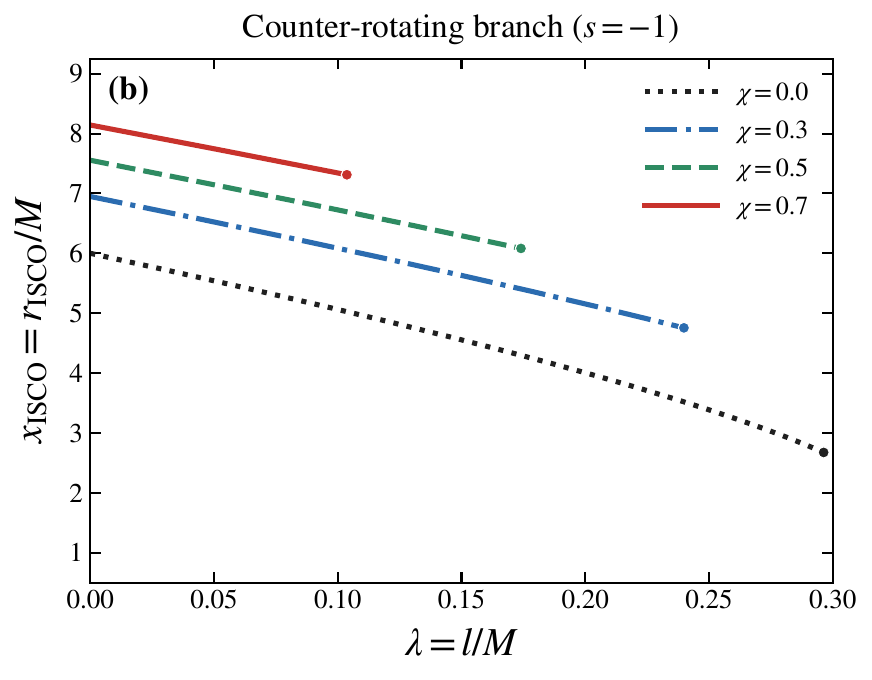}

    \caption{
    Dimensionless ISCO radius
    \(x_{\mathrm{ISCO}}=r_{\mathrm{ISCO}}/M\) as a function of the
    Fan--Wang charge parameter \(\lambda=l/M\) for fixed rotation
    parameters \(\chi=a/M=0\), \(0.3\), \(0.5\), and \(0.7\).
    Panel (a) shows the co-rotating branch \(s=+1\), while panel (b)
    shows the counter-rotating branch \(s=-1\).
    The curves represent the physical ISCO solutions outside the
    outer event horizon, while their terminal markers indicate the
    limiting extremal configurations. The curves are not extended
    beyond these endpoints because no event horizon exists in that
    region of the parameter space. At \(\lambda=0\), the corresponding
    Kerr ISCO radii are recovered.
    }
    \label{fig:fw_isco_charge}
\end{figure*}

As a consistency check, in the limit \(\lambda=0\), the ISCO equation reduces to the standard Kerr condition\cite{chandrasekhar1998mathematical,chaverri2025isco}
\begin{equation}
\label{kerr_isco}
x^2-6x+8s\chi\sqrt{x}-3\chi^2=0.
\end{equation}
For \(\chi=0\), it reduces to
\begin{equation}
\label{P_func_0}
P(x,\lambda)=0,
\end{equation}
which is precisely the ISCO equation of the static Fan–Wang spacetime.
The combined influence of the Fan–Wang charge parameter and rotation is illustrated in Fig.~\ref{fig:fw_isco_charge}, where the dimensionless ISCO radius is shown as a function of \(\lambda\) for several fixed values of \(\chi\).

Within the BH domain, increasing \(\lambda\) shifts both the co-rotating and counter-rotating ISCOs toward smaller radii. This displacement is substantially stronger for the co-rotating branch. At \(\lambda=0\), all curves reproduce their corresponding Kerr values. For each fixed value of \(\chi\), the curve terminates at the spin-dependent extremal boundary determined by the simultaneous conditions for a degenerate horizon. Beyond this boundary, the spacetime has no outer event horizon and is therefore not included in the present BH analysis.

\section{Data analysis and results}
\label{sec:defangle}

To assess the viability of the Fan-Wang spacetime in describing BHs in X-ray binaries, we analyze $(f_U^o,f_L^o)$ pairs of QPO frequencies observed in four selected microquasars: GRO~J1655-40 \cite{Motta:2013wga}, GRS~1915+105 \cite{2006ApJ...637.1002R}, H1743-322  \cite{Molla:2016mip,2014MNRAS.444.2065I}, and XTE~J1550-564 \cite{Remillard:2002cy,2011ApJ...730...75O}.

\begin{table*}
\centering
\footnotesize
\setlength{\tabcolsep}{.62em}
\renewcommand{\arraystretch}{1.4}
\begin{tabular}{lccccccccccc}
\hline\hline 
Microquasar
& \multicolumn{2}{c}{QPO pairs}
& \multicolumn{4}{c}{Priors: Gaussian $\&$ uniform}
& \multicolumn{4}{c}{Best-fit parameters}\\
& $f_U^{\rm o} [Hz]$ & $f_L^{\rm o} [Hz]$ & $M [M_{\odot}]$ & $a/M$ & $r/M$ & $l/M$ & $M [M_{\odot}]$ & $a/M$ & $r/M$ & $l/M$ \\
\hline  
GRO J1655-40                &
$441^{+2.0}_{-2.0}$         &
$298^{+4.0}_{-4.0}$         &
$(5.4;0.3)$                 &
$(0;2/3)$                   &
$(5.7;0.2)$                 &
$[-3,+3]$                   &
$5.41^{+0.21}_{-0.21}$      &
$0.40^{+0.37}_{-0.34}$      &
$5.66^{+0.17}_{-0.18}$      &
$-0.04^{+0.14}_{-0.15}$     \\
GRS 1915+105                &
$168^{+3.0}_{-3.0}$         &
$113^{+5.0}_{-5.0}$         & 
$(12.4;1.9)$                &
$(0;2/3)$                   &
$(6.2;0.6)$                 &
$[-3,+3]$                   &
$12.48^{+1.29}_{-1.29}$     &
$0.17^{+0.60}_{-0.59}$      &
$6.15^{+0.33}_{-0.36}$      &
$-0.01^{+0.25}_{-0.24}$     \\
H1743-322                   &
$240^{+3.0}_{-3.0}$         &
$165^{+9.0}_{-5.0}$         & 
$(11.2;1.8)$                &
$(0;2/3)$                   &
$(5.3;0.6)$                 &
$[-3,+3]$                   &
$11.32^{+1.21}_{-1.22}$     &
$0.29^{+0.55}_{-0.53}$      &
$5.01^{+0.28}_{-0.31}$      &
$+0.05^{+0.22}_{-0.22}$      \\
XTE J1550-564               &
$276^{+3.0}_{-3.0}$         &
$184^{+5.0}_{-5.0}$         & 
$(9.1;0.6)$                 &
$(0;2/3)$                   &
$(5.5;0.2)$                 &
$[-3,+3]$                   &
$9.12^{+0.42}_{-0.42}$      &
$0.46^{+0.43}_{-0.39}$      &
$5.43^{+0.19}_{-0.20}$      &
$-0.03^{+0.16}_{-0.16}$      \\
\hline
\end{tabular}
\caption{MCMC results for the four selected microquasars. Columns list, respectively: source name, observed upper and lower frequencies, Gaussian priors $(\mu_i;\sigma_i)$ on the masses $M$ (in solar masses), angular momenta $a/M$, and radii $r/M$, uniform priors on the magnetic charges $l/M$, and the best-fit values of $M$, $a/M$, $r/M$, and $l/M$.}
\label{tab:1}
\end{table*}

The best-fit parameters of each source are deduced via MCMC analyses and the parameter space is sampled via the  Metropolis-Hastings algorithm. 
The posterior probability is given by
\begin{equation}
\label{bayes}
P(\theta|D,I) = \frac{\mathcal L(D|\theta,I) \mathcal{P}(\theta|I)}{P(D|I)} \, ,
\end{equation}
where $\theta$ is the model parameter vector, $D$ is the observational dataset, $I$ is the set of assumptions, $P(\theta|D,I) $ is the posterior probability, $\mathcal L(D|\theta,I) $ is the likelihood function, $\mathcal{P}(\theta|I)$ is the prior probability, and $P(D|I)$ is the normalization.

Since each microquasar has only one observed QPO pair, to better constrain the model parameters, we set the following prior probabilities $\mathcal{P}(\theta|I)$:
\begin{itemize}
\item[-] Gaussian priors on: a) the masses $M$, deduced from observations \cite{Motta:2013wga,Remillard:2004sp,2014MNRAS.439L..65M,Ingram:2014ara,2011ApJ...730...75O,Xamidov:2025oqx}, b) the angular momenta $a$, within the condition for maximally rotating BH in the Kerr spacetime, i.e. $|a|\leq M$, and c) the radii $r/M=(M f_U^{\rm o})^{-2/3}$, based on the Schwarzschild spacetime relation, where $f_U^{\rm o}$ are the observed upper frequencies;
\item[-] uniform priors for the charge parameter $l/M$.
\end{itemize}
In view of the above considerations, we have
\begin{equation}
\mathcal{P}(\theta,I)\!\propto\!\left\{\!
\begin{array}{ll}
\!\displaystyle \exp\!\left[\!-\frac{(\theta\!-\!\mu_i)^2}{2\sigma_i^2}\!\right]\!, & \theta\!=\!\{M,r/M,a/M\},\\
\!1, & \theta = l/M.
\end{array}
\right.
\end{equation}
where mean values $\mu_i$ and variances $\sigma_i$ of the Gaussian priors and the uniform priors are summarized in Table~\ref{tab:1}.

\begin{figure*}[htbp]
\centering
\includegraphics[width=0.75\hsize,clip]
{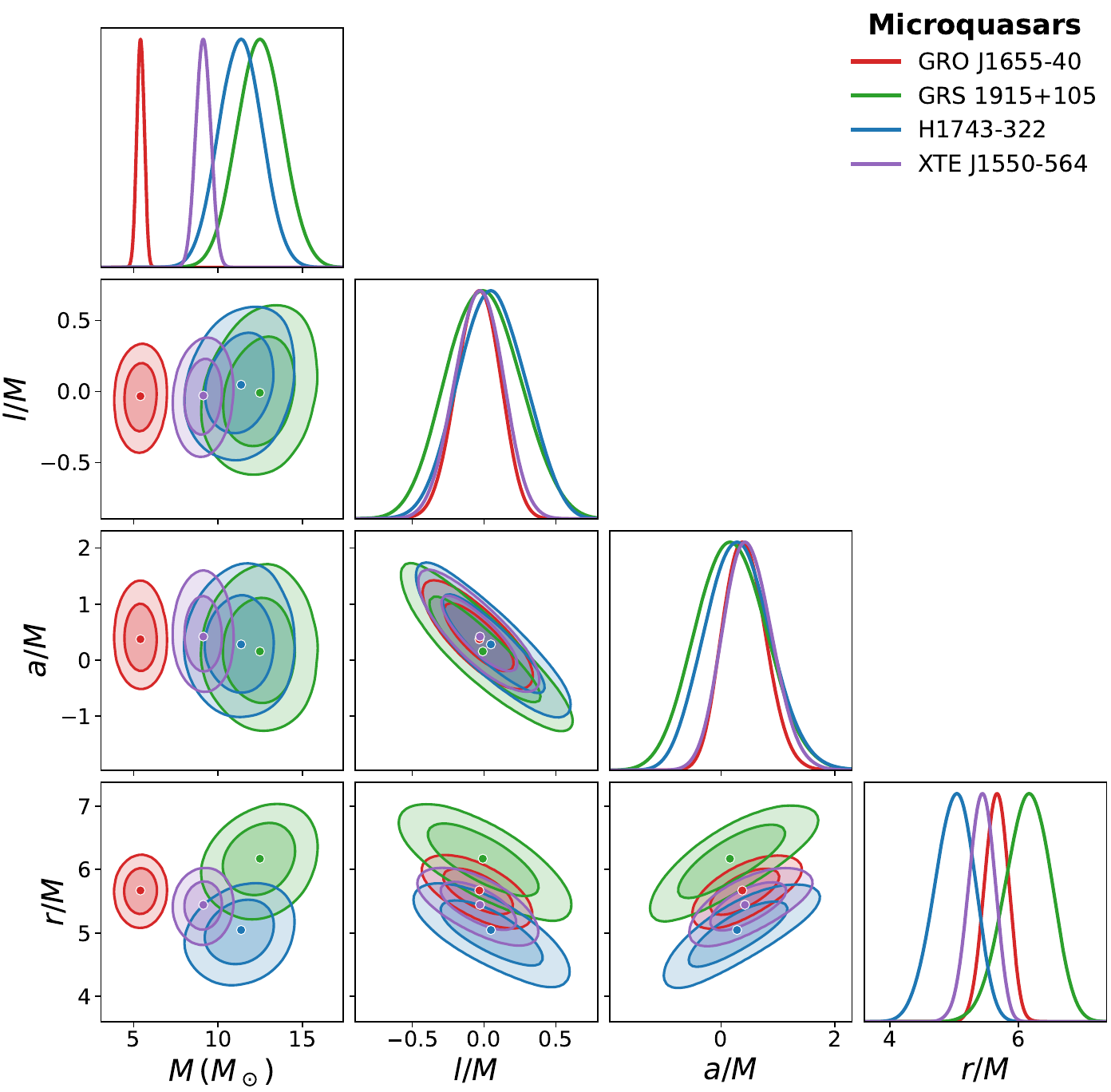}
\caption{MCMC posteriors on the parameters $\{M,\,l/M,\,a/M,\,r/M\}$ obtained from selected microquasars. The dark (light) shaded regions correspond to $1\sigma$ ($2\sigma$) confidence levels, whereas the dots correspond to the best-fit values listed in Table~\ref{tab:1}.}
\label{fig:contours}
\end{figure*}

For each source, the MCMC analysis aims at maximizing the log-likelihood function in Eq.~\eqref{bayes} defined as
\begin{equation}
\label{loglike}
\ln \mathcal L = \ln \mathcal L_U + \ln \mathcal L_L\,,
\end{equation}
which incorporates data $(f_U^{\rm o},f_L^{\rm o})$ and errors $(\sigma_{f_U^{\rm o}},\sigma_{f_L^{\rm o}})$ from upper and lower frequencies as follows
\begin{equation}
\ln \mathcal L_i = - \frac{(f_i - f^{\rm o}_i)^2}{2\sigma_{f_i^{\rm o}}^2} - \ln\sqrt{2\pi \sigma_{f_i^{\rm o}}^2}
\quad,\quad
i=\{U,L\}.
\end{equation}
The corresponding best-fit results are summarized in Table~\ref{tab:1} and the MCMC posteriors are portrayed in Fig.~\ref{fig:contours}.

The fit performed in an extended parameter domain: negative $l/M$ values are allowed only to locate the best-fitting frequencies and are not interpreted as physical charges. The Fan-Wang RBHs
correspond to $l \ge 0$.

As described in Sec.\ref{sec:gravred} the fit uses
$f_{U}=f_{\phi}$ and $f_{L}=f_{\phi}-f_{r}$; the nodal-precession frequency is not
included. The mass estimates remain close to the observational values adopted as priors. This agreement should be read in light of the prior information, since each source supplies only one pair of HF QPOs. The fitted emission radii, $r/M \simeq 5.0-6.2$, place the oscillations in the inner disk region. For all four sources, $l = 0$ lies within the reported $1\sigma$ interval. Thus the extended RPM fit reveals no statistically resolved departure
from the Kerr limit. The negative central values in Table ~\ref{tab:1} are fitting results, not evidence for negative
physical charges.

Because the metric reduces to Kerr for $l=0$, published Kerr RPM
estimates provide a useful reference (Table~\ref{tab:kerr_comparison}).
For GRO~J1655-40, our mass $M=(5.41\pm0.21)\,M_\odot$ and radius
$r/M=5.66$ agree with $(5.31\pm0.07)\,M_\odot$ and $5.68\pm0.04$
from Ref.~\citep{Motta:2013wga}. For XTE~J1550-564, our radius $5.43$ agrees
with $5.47\pm0.12$ \cite{Ingram:2014ara}. Both sources have broad spin intervals that
include the Kerr values listed in Table~\ref{tab:kerr_comparison}.

For GRS~1915+105, our mass $(12.48\pm1.29)\,M_\odot$ is broadly
compatible with the Kerr-limit value $13.99\,M_\odot$~\cite{banerjee2023signatures}, and our spin interval includes the corresponding value $\simeq 0.3$.
For H1743-322, our central mass and spin satisfy the bounds in Ref.~\cite{Ingram:2014ara}. Its Kerr mass estimate in Ref.\cite{yang2025observational}, $9.745\,M_\odot$, is lower than our central value, whereas the spin $0.283\pm0.003$ is close to ours. The much larger uncertainties of the present
fit and the different inputs must be considered in interpreting
this comparison.

The smaller spin errors of the published Kerr analyses cannot be compared solely as a metric effect. For
GRO J1655-40, \cite{Motta:2013wga} uses three simultaneous QPOs, including the low-frequency signal identified with
nodal precession. For XTE J1550-564, the result quoted from Ref. \cite{Ingram:2014ara} instead uses a low-frequency QPO, a
lower HF QPO, and an independently estimated mass. The H1743-322 bounds in that reference use the HF
pair together with the highest observed type-C QPO frequency and an ISCO condition. Our fit uses only the
twin HF QPOs and contains the additional parameter $l$. Table ~\ref{tab:kerr_comparison} therefore provides a literature comparison, rather than a controlled comparison based on identical data and priors.

Figure~\ref{fig:contours} also shows a pronounced anticorrelation
between $a/M$ and $l/M$. At fixed mass and radius, increasing positive
$l$ reduces the Fan-Wang mass function and changes the orbital and
precession frequencies. Changes in spin and emission radius can
partly compensate for this effect, contributing to the broad
intervals. Adding a simultaneous nodal-precession measurement
would supply another constraint; such joint QPO analyses have
been applied to other rotating regular and charged BH metrics \cite{wu2026using,ali2026precessions}.

For the same Fan-Wang metric, EHT shadow analyses give
$l/M\lesssim0.08$ for M87$^*$ at inclination $17^\circ$ and
$l/M\lesssim0.03$ for Sgr~A$^*$, under the assumptions of Ref.~\cite{2026PDU....5102203G}.
Our uncertainties have a larger characteristic scale. This is a
qualitative comparison because the quoted shadow bounds and our
extended fit have different statistical constructions. QPOs
nevertheless provide a complementary probe of massive-particle
motion around stellar mass BHs, whereas the shadow
constraints probe photon motion around supermassive objects.

\begin{table*}[t]
    \centering
    \caption{Comparison with published Kerr RPM estimates.
    The Fan-Wang columns refer to the extended fit of
    Table~\ref{tab:1}. The literature entries use different
    observational inputs and fitting procedures.}
    \label{tab:kerr_comparison}

    \begin{ruledtabular}
    \begin{tabular}{lcccc}
        Source
        & \shortstack{This work:\\$M\,[M_\odot]$}
        & \shortstack{This work:\\$a/M$}
        & \shortstack{Kerr:\\$M\,[M_\odot]$}
        & \shortstack{Kerr:\\$a/M$} \\
        \hline

        GRO~J1655-40
        & $5.41\pm0.21$
        & $0.40^{+0.37}_{-0.34}$
        & $5.31\pm0.07$
        & $0.290\pm0.003$ \\

        & & &
        $5.307^{+0.024}_{-0.021}$
        & $0.287\pm0.002$ \\

        GRS~1915+105
        & $12.48\pm1.29$
        & $0.17^{+0.60}_{-0.59}$
        & $13.99$
        & $\simeq 0.3$ \\

        H1743-322
        & $11.32^{+1.21}_{-1.22}$
        & $0.29^{+0.55}_{-0.53}$
        & $\geq 9.29$
        & $\geq 0.21$ \\

        & & &
        $9.745^{+0.092}_{-0.082}$
        & $0.283\pm0.003$ \\

        XTE~J1550-564
        & $9.12\pm0.42$
        & $0.46^{+0.43}_{-0.39}$
        & \shortstack{$9.1\pm0.6$\\(mass input)}
        & $0.339\pm0.007$ \\
    \end{tabular}
    \end{ruledtabular}

    \par\smallskip
    \begin{minipage}{\linewidth}
        \footnotesize
        For Ref.\cite{banerjee2023signatures}, the listed values are the profiled $\chi^2$
        minimum for the RPM in the Kerr limit of the Bardeen model
        (charge parameter $g\simeq 0$), rather than an independent
        mass measurement. For XTE~J1550-564, the literature mass
        was an external input, not a QPO-derived output.
    \end{minipage}
\end{table*}

\section{Conclusions}
\label{sec:con}

In this work, we investigated the orbital dynamics and QPOs in the rotating Fan--Wang spacetime. Starting from the static Fan--Wang solution, we considered its stationary and axisymmetric extension constructed through the modified Newman--Janis procedure. The resulting geometry recovers the static Fan--Wang spacetime in the nonrotating limit and reduces to the Kerr solution when the magnetic charge parameter vanishes. We analyzed its horizon and stationary-limit structure and obtained the extremality conditions that determine the BH domain in the parameter space.

For neutral massive particles moving along equatorial circular orbits, we derived the Keplerian angular velocity, specific energy, and specific angular momentum for both co-rotating and counter-rotating motion. Small perturbations about these orbits were then used to obtain the radial and vertical epicyclic frequencies. The corresponding expressions reproduce the static Fan--Wang results for $a\to0$ and the Kerr expressions for $l\to0$. The ISCO condition was formulated in terms of the radial epicyclic frequency and solved numerically. Rotation separates the co-rotating and counter-rotating branches, moving the former inward and the latter outward. For a fixed spin, increasing the Fan--Wang parameter shifts both ISCO branches toward smaller radii, with a stronger effect on the co-rotating branch.

The derived frequencies were applied within the RPM to the observed twin HF QPOs of GRO J1655--40, GRS 1915+105, H1743--322, and XTE J1550--564. The MCMC analysis provided simultaneous estimates of the BH mass, spin, Fan--Wang parameter, and QPO emission radius. The inferred masses remain close to the observational estimates adopted as priors, while the emission radii, $r/M\simeq5.0$--$6.2$, locate the oscillations in the inner regions of the accretion disks. The resulting mass, spin, and radius estimates are broadly compatible with published Kerr RPM analyses, although the latter employ different observational inputs and fitting assumptions.

For all four sources, the Kerr limit $l=0$ lies within the reported $1\sigma$ intervals. The present QPO data therefore do not provide statistically significant evidence for a nonzero Fan--Wang parameter. The posterior distributions also reveal a pronounced anticorrelation between $a/M$ and $l/M$, showing that changes in the charge parameter can be partially compensated by the spin and emission radius. The negative central values obtained for some sources result from extending the sampled parameter domain around $l=0$ and should not be interpreted as physical negative charges.

The current constraints are limited by the use of only one pair of HF QPOs for each source and by the simultaneous fitting of four model parameters. A future analysis incorporating a low-frequency QPO associated with nodal precession, together with independent measurements of mass and spin, could reduce the parameter degeneracies and provide a more stringent test of the Fan--Wang geometry. The results obtained here establish a framework for confronting the rotating Fan--Wang spacetime with X-ray timing observations and for distinguishing its strong-field orbital properties from those of the Kerr spacetime.

\section*{Acknowledgment}
This research is funded by the Abai Kazakh National Pedagogical University (Contract No  05-04/568. dated 21.05.2026)

\bibliography{0refs}
\end{document}